\documentclass[runningheads]{llncs}
\usepackage{graphicx}
\begin{document}
\title{Lightweight Mobile Ad-hoc Network
Routing Protocols for Smartphones}
\titlerunning{Lightweight MANET routing protocols}
%
\author{Md Shahzamal}
%
%
\institute{Macquarie University, Sydney, Australia\\
\email{md.shahzamal@students.mq.edu.au}}
\maketitle              
\begin{abstract}
Mobile Ad-hoc Network (MANET) is one of the promising wireless networking approaches where
a group of wireless devices can establish communication among themselves without any
infrastructure. MANET has successfully been implemented in the emergency communication,
battlefield and Vehicular Ad-hoc Network (VANET) etc. Recently, MANET has also been studied to
establish village level off-the-grid telephony system using mobile phones (smartphones). As the
mobile phones and other modern electronic devices are increasingly equipped with efficient
wireless Wi-Fi devices, phone based MANET implementation will open new horizons for off-the-self
services for phones as well as IoT implementations. However, current MANET routing protocols show poor performances in implementing MANET among mobile phones due to the requirements of high memory, commutation power and
high energy usage. The performance seriously degrades when the number of phones
increases in the network. Therefore, a lightweight routing protocol is the key requirement for
successful implementation of MANET using mobile phones. This report presents relevant routing
protocol designs and their applicability to develop lightweight MANET routing protocols for mobile
phones. In conclusion, we have presented the challenges and research directions in this field.

\end{abstract}
\section{Introduction}
Mobile Ad-hoc Network (MANET) is a self-reconfigurable and decentralize communication system
that consists of a collection of wireless devices (nodes). These wireless nodes can move
independently while participating in the network. In the MANET, the nodes not only act as users but
also as routers to forward data. Therefore, a source node can send data to the destination node
directly or through intermediate nodes when the destination node is not within the communication
range~\cite{roy2010handbook}. As MANET does not depend on any fixed stations, the network can be formed anywhere
at any time. In the deployment of network, available wireless devices such as laptops, PDAs and
phones etc. are the main communication resources for executing network operations along with
other few wireless networking devices. As a result, the implementation cost is quite low in
comparison with traditional infrastructure based communication systems. Due to such flexibility and
potentiality, MANET has intensively been studied to implement networks in the extreme
environments where network infrastructure building is impossible or traditional networking is cost
prohibitive according to application objectives.

MANETs are widely implemented as the emergency communication system in the disaster area
where quick deployment of network is highly anticipated to support recovery operations~\cite{bai2010emergency}. Military
operation is another scenario where MANET has shown its potentiality to deploy tactical
communication network that connect soldier hand-held devices, mobile tanks, distributed sensors
and aircrafts~\cite{burbank2006key}. Except these temporary deployments, MANET is also applied in the Vehicular
Area Network (VANETs) for traffic management and route conditions dissemination. Recently,
MANET has been investigated to establish village level telephony system using mobile phones.
This telephony system provides cost free services such as text messaging and VOIP call among
villagers without traditional network~\cite{adeyeye2011village,gardner2013serval}. This is being considered as an alternative communication
system for rural area in developing countries. Mobile phone based MANET (Phone-MANET) has
created a new horizon for developing off-the-self services of phones. These services include
content sharing among a community, messages dissemination in the conferences, convention
centres and shopping malls and lecture distribution in the classroom etc. [25]. As this is the off-thegrid communication among mobile phones, the services are cost free and reduce the traffic load for
infrastructure based network. Phone-MANET could also be combined with the forthcoming Internet
of Things (IoT) and hence improves the services of IoT.

Although information sharing among mobile phones through direct communication is not new. This
peer-to-peer (P2P) communication service is available with mobile phones via Bluetooth
technology for long time. Mobile phones can share photos or videos using this technology over
short-communication range. Current Bluetooth technology can maintain file sharing among a group
of seven devices~\cite{jung2007bluetorrent}. 802.11 Wi-Fi can also be used to implement peer-to-peer sharing among
phones as the Wi-Fi standard supports ad-hoc mode along with infrastructure mode. The Wi-Fi
technology offers much higher data speed and communication range than Bluetooth. However, WiFi enabled phones can only make ad-hoc network within one hop. The modern electronic devices
are increasingly equipped with Wi-Fi devices and communication among these devices is highly
desired in many applications. But, 802.11 Wi-Fi cannot establish efficient device-to-device
communication. Due to this lack of existing solutions, a new approach of Wi-Fi called Wi-Fi Direct
has been proposed. This technology operates communication among a group of phones mimicking
the approach of access point (AP) based communication. In a group, a device takes over the role
of AP that is called P2P Group Owner (P2P GO), while other devices are P2P clients which
communicate through P2P GO. The Wi-Fi Direct enabled devices can act as both P2P GO and
P2P client. One group can connect with other group being client to that group. However, the P2P
Clients cannot connect more than one group and P2P client cannot be P2P GO for its GO~\cite{camps2013device}. This
prohibits scalability of the network. As the Wi-Fi Direct enabled devices cannot be client and router
for each other, MANET routing protocol cannot function on Wi-Fi Direct. Besides, Wi-Fi Direct is
required to configure manually. Therefore, it cannot implement self-organised network~\cite{zheng2014hybrid}. As the
peer-to-peer communication among mobile phones would play a vital role in the future communication arena, researchers have moved to make a generalised solution. The research has drifted to modify the 802.11 Wi-Fi radio available with phones for multi-hop ad-hoc communication.
As many smartphone manufacturers have open the source codes of operating system, multi-hop
can be achieved by modifying the Wi-Fi driver.

The authors of~\cite{spart,chowdhury2013olsr} have modified the Wi-Fi radio of smartphones and implemented multi-hop
ad-hoc mode within. With these developments, phone-to-phone communication became fully selforganized and MANET routing protocol can be implemented. However, the network size still
remains very small. The reason is that current MANET routing protocols require huge computing
power, energy and memory that phones cannot provide. The previous phone-MANET base
projects have adopted the popular MANET routing protocol called Optimised Link State Routing
Protocol (OLSR). OLSR works pro-actively where every node maintains full routing paths to every
destination in the network. Nodes achieve this global information maintenance by periodic flooding
of link information about its neighbours. Route reconstruction is done continuously based on this
flooded information. In this protocol, phones are required to store information about the entire
network topology when they take part in MANET. This is the main bottleneck to implement MANET
over phones. When the number of nodes increases the network performance degrades sharply.
Besides, this protocol drains phones energy quickly as this protocol generates huge control
overhead. The control overhead is generated as any change in the network is flooded into the
entire network. Therefore, to make successful MANET application using mobile phones, protocol
should be such that phones are not required to store the whole network topology to establish paths
to the destination and control overhead would be limited.

In this study, we have presented some popular MANET routing protocols that can be considered to
address lightweight routing protocol for smartphones. In the section-II, we have described the
state-of-the-arts of relevant protocols and section-III has presented the open issues in this field.
We have concluded our study in the Section-IV.

\section{Approaches to Lightweight Protocols}
In the context of implementing MANET over smartphones, lightweight routing protocols refer those
protocols that require minimal memory for routing table, less computing resources as well as
generating less protocol control overhead. Reactive routing protocols which initiate route finding on
demand do not maintain routing table. These protocols can provide lightweight routing protocols,
but their convergence time is high and often create loops in paths~\cite{hong2002scalable}. Therefore, the proactive
protocols that do not depend on global network information or hybrid protocols that combine
proactive and reactive protocols can be good candidate. In the following paragraphs, we have
described the relevant methods.

\subsection{BATMAN Routing Protocol}
Better Approach to Mobile Ad hoc Network (BATMAN) is the routing protocol designed for MANET
that does not maintain full path to the destination. Each node only collects and maintains the
information about the best next hop towards all other nodes in the network. Nodes collect this
information through hello packets, known as originator messages (OGMs), broadcasting by every
node periodically. The OGMs at least contain the information about the originator address, sender
address and a sequence number. Upon receiving a new OGM from neighbours, a node checks the
sequence number and rebroadcasts OGM replacing its own address to the sender address if the
message was not received before. The node also keeps track of which neighbour has sent the
maximum number of OGM messages from an originator for a period called window time. The
respective neighbour is considered as the best next hop to the originator node. When a node has
data packet to send to a given destination, it will forward the packet to the best neighbour who
forwarded maximum OGM messages generated by the destination node. The process is repeated
in the next hops and data packet is delivered to the final destination. This protocol removes link for
a node if the OGM message from that node is not received for a time-out period~\cite{neumann2008better}. As the
protocol does not keep complete routing paths to the destinations, the protocol is suitable for
storage constrained devices. Best next hop selection over a window time and validation of links for
a time-out period avoid routing loops. Since this protocol depends only on hello packets and does
not broadcast topology change messages, the control overhead is low. However, this protocol also
degrades performance when the network size grows and nodes mobility increase. This is because
nodes cannot be updated accurately and quickly about other nodes. Therefore, the packet drop
increases as it cannot find the destination node. The situation is improved by the authors of~\cite{sujatha2015uoshr}
who have proposed to add a reactive process when the packets do not find the next hop at the
source or in the path to the destination. Reactive process broadcasts request RREQ within a zone
of node where a packet does not find the next hop to check which nodes have the next hop to the
destination. Then the protocol forwards the packet to that (which has the next hop). This
modification improved the performance of BATMAN protocol. However, there are more chances to
set up infinity forwarding states as rebuilding next hop cannot guarantee the delivery of packet to
the destination.

\subsection{Zone Based Routing Protocols}
In the zone based routing protocols (ZRPs), each node is only required to know the nodes that are
within a specified hop distance. This property is highly desired for memory constrained devices. In
these protocols, a zone is created with the centre at the node and a radius of specified hop
distance. Each node maintains paths to every node within its zone through an Intra-zone Routing
Protocol (IARP). This IARP could be any proactive routing protocol. When the destination node is
outside the zone, the path set-up is reactive. To find the routes outside the zone, node first sends a
route query to the boarder nodes on the periphery of its zone. If the border nodes find the destination nodes in their won zones, they send back a route replay on the reverse path. Otherwise, it rebroadcasts the route query to their border nodes again and the process continues until the destination is found. The broadcasting from border to border is called Bordercasting. The
global route discovery is done using Inter-zone Routing Protocol (IERP) and Bordercast Resolution
Protocol (BRP). IERP is usually the enhanced version of reactive protocols that can discover and
maintain routes depending on the link information provided by IARP. BRP also utilises topology
information of IARP to direct query request to the border of the zones. BRP employs a query
control procedure that avoids rebroadcasting of query to the areas of network that have already
been covered~\cite{beijar2002zone}. This protocol reduces the control overhead of proactive routing protocols as
well as path finding latency of reactive protocols.

Several protocols have been developed based on this basic approach. AOHR (AODV and OLSR
Hybrid Routing) is one of the ZRP based routing protocols where nodes use OLSR to set up paths
inside the zone and a modification of AODV for outside the zone~\cite{shaochuan2006aohr}. Zone radius is selected
dynamically by AOHR to be applicable in various network scenarios. The modified AODV uses
Multipoint Relaying (MPR) for boarder-casting the global route query. This reduces the control
overhead than the basic ZRP. This protocol is lightweight in terms of memory requirements and
overhead control. The AODV rediscovers the route newly if the path is broken during data
transmission. Besides, the routing path to the destination outside the zone is not the shortest path
as the paths are created through border nodes of traversed zones. This decreases the network
throughput and packet delivery ratio. The situation becomes worse for the dynamic networks. The
authors of~\cite{kumar2008genetic} have proposed the Genetic Zone Routing Protocol (GZRP) that addressed this
limitation of AOHR. The border nodes maintain multiple optimal or sub-optimal paths for the
destination node inside the zones. These paths are found by a genetic algorithm that borders node
apply on the topological database available with them.

\subsection{Cluster Based Routing Protocols}
Clustering is another approach to divide the network into groups of nodes that are geographically
close. Each cluster selects a head that coordinates its activities. The cluster also selects some
nodes on the periphery of cluster as gateways that are responsible for intercommunicating among
clusters. For establishing communication outside the cluster, cluster heads can communicate with
each others~\cite{jiang1999cluster}. In the zone based routing protocols, the zones are overlapping whereas the
clusters are separated from each other. Therefore, control messages may traverse over more than
one zone in ZRP. On the other hand, control messages are restricted within the cluster and hence
clustering techniques could reduce more routing control overheads. The performance of the cluster
based routing protocols depends not only route discovery and route maintenance but also on the
cluster formation mechanism. The route discovery is done by using either proactive or hybrid
methods that combine proactive and reactive processes. Cluster formation is done using tree formation, maximum node degree based head selection and hop-distance based clustering etc. As
the protocols can operate with the partial view of network, this provides the possibility to develop
efficient MANET routing protocols for smart-phones. The following paragraphs have descried the
popular cluster based routing protocols.

Clustering based on the tree formation is a popular way of grouping nodes in the MANET. The
authors of~\cite{baccelli2006olsr} have made clustering based on the tree formation where roots of the trees are
cluster heads and the leaf nodes that have neighbours from other clusters are the gateway nodes.
The plain OLSR protocol is applied inside the cluster to find the path. Another upper level OLSR is
applied among the cluster heads. The cluster heads know the link state information that other
nodes have. The cluster heads maintain the paths to the nodes that are in other cluster. For
communicating with the nodes outside the cluster, node sends data through cluster heads. The
size of routing table for cluster members becomes smaller and routing control overhead is also
reduced as topology change broadcasting is restricted within the cluster. However, cluster heads
require more storage and computing power as well as tree maintenance produce overhead. This
approach is suitable for heterogeneous network where there are some resourceful devices. The
authors of~\cite{ros2007cluster} have proposed protocol called Cluster-OLSR (COLSR) that is independent of
underlying clustering algorithm. Another difference from the previous one is that nodes send data
packets to the respective cluster head for outside communication and then cluster head forwards
the data packet to the destination node. In this protocol cluster head is only required to maintain
path to other cluster heads. Thus, cluster heads require less resources. In these protocols, cluster
heads might have transmission loads to maintain routes for other clusters. To balance the load of
cluster heads, the authors of~\cite{kannhavong2008sa} have proposed SA-OLSR that broadcasts the Cluster TC
messages over the entire network. The other clusters receive only the first copy of Cluster TC and
assess the traversed path by the first copy as the faster and less congested path while other
copies are discarded. As the route discovery follows the less congested path and the load of
cluster heads becomes balance.

The above cluster based routing protocols use proactive route discovery inside the cluster as well
as outside the cluster. These methods provide faster route establishment between source and
destination. However, the cluster heads should be special node for extra responsibility. Therefore,
routing performance decreases for homogeneous network like phone-MANET. If the combination of
proactive and reactive approach is applied for cluster based routing protocol, cluster heads do not
require to store much information. The authors of~\cite{sharma2013efficient} have proposed a cluster based hybrid
routing protocol. In this protocol, the intra-cluster communication is done using proactive routing
protocol (DSDV) and inter-cluster routing is done using source routing based reactive routing
protocol. When the destination node is not inside the cluster, node sends a route RREQ messages
to every cluster through cluster heads. The cluster head that has the destination node sends back RREP through the gateways and cluster heads. In this protocol, clusters do not keep information
about the nodes that belongs to other clusters, except its own cluster member nodes. Therefore,
requirements of special nodes for cluster head is diminished with the cost of global path finding
latency. However, the cluster head is still responsible for maintaining the source to destination
information. The authors of~\cite{huang2009hybrid} have proposed a method to get back RREP through the different
paths instead of the paths that has followed by RREQ. This reduces the load of cluster head and
the protocol becomes more applicable for homogeneous network.

\subsection{Link-State Routing Protocols}
The most popular link-state routing protocol is the Optimized Link State Routing protocol
(OLSR)~\cite{clausen2003optimized}. This protocol maintains routes to every nodes in the network. This is done by flooding
the link-state information throughout the entire network. OLSR uses an optimization technique to
flood the link-state information. This protocol broadcasts a subset of link instead of all links called
multipoint relay selector. This subset of links is flooded through the Multipoint Relaying and this
reduces the redundant transmissions in the network. These properties make the OSLR more
stable than other link-state routing protocols. However, due to its larger routing table size this is not
efficient to deploy MANET using memory constrained devices like smartphones. The OLSR is
designed for all kind of communication. After tuning some parameters, this can be used to
implement network with more nodes. Besides, stable MPR selection will produce less topology
control overhead. This could be beneficial for energy constrained devices.

The authors of~\cite{guizani2012new} have proposed a Cluster-Based Link State Routing Protocol (CLSR) that works
pro-actively. This protocol does not produce any cluster formation and maintenance mechanism as
the clustering is done using routing information. This protocol uses hello messages to forms onehop cluster depending on the node connectivity. The nodes who have the maximum of neighbours
becomes head and inform other through the next hello messages and the member nodes inform its
neighbour about its cluster head. This protocol uses another message called CTC (Cluster
Topology Control) to discover neighbour clusters. The cluster heads are connected using
Connected Dominating Set (CDS) to form a virtual mesh backbone. Cluster topology information is
broadcasted through the gateways. The cluster heads make the routing table only on behalf of the
cluster members. That reduces the routing table size.

\section{Challenges and Future Directions}
Mobile Ad hoc Networking (MANET) among smartphones is the new dimension of wireless
communication. Researchers have already managed to integrate multi-hop ad-hoc communication
mode with the 802.11 Wi-Fi radio of smartphones~\cite{jung2014designing}. However, mentionable research has not
been done to develop appropriate routing protocols for implementing MANET using mobile phones
in the bigger domain. Most of the previous smartphones based MANET projects~\cite{spart,jung2014designing} have applied OLSR and can only manage network for few users. In this report, we have presented some
relevant protocols. Different protocols have different prospects for implementing phone-MANET
over various network environments. BATMAN is one of the simplest MANET routing protocol. This
protocol just maintains the best next hop to every destination and produces very low control
overhead although this works pro-actively. This has the characteristics to be applicable for phoneMANET routing protocols. However, this protocol is not much scalable. This is because when the
network size grows, the OGM messages are required more time to notice other nodes about its
presence. The nodes which do not get updated within the time-out period will delete routing
information to the nodes. The mobility of nodes also affects the performance of this protocol
significantly. As the forwarded packets will lose the way if the node changes its position quickly in
the horizontal direction. The modification of BATMAN [20] can improve for certain network size.
After that the protocol becomes fully reactive and loop creations will occur more often due to the
internal mechanism of BATMAN. If the directional information of movements is combined with the
best next hop selection, the performance could be improved. The BATMAN protocol will outperform
other protocols where the node movements are low and MANET services are not real-time. The
BATMAN protocol can also be applied inside the cluster instead of OLSR in cluster based routing
protocol that would reduce routing table size and control overheads significantly.

Network segmentation is a good way to reduce storage requirements for routing. However, in the
ZRP each node is required to run three protocols (IARP, IERP and BRP) for network operation.
That will increase computational load as node number grows. As the network size goes up, the
zone radius increases too, optimisation by Bordercasting protocol will also reduce as the target
border nodes will shift position and became border nodes of other zones. This also increases the
uncertainty of finding destination nodes. The situation can be controlled keeping the radius low.
However, the latency of path finding will increase significantly as well as will increase the control
overhead. The zone routing protocol does not maintain shortest paths between source and
destination. The authors of [22] have proposed to use genetic algorithm (GA). But, GA usually
requires complex computation that could not afford the smartphones. Simple alternative path
maintenance procedure is required for the improvement of ZRP routing protocols. This protocol is
suitable for the environment where all participating nodes have the same communication capability.
Other network partitioning method is clustering that requires special nodes as cluster heads when
the protocol is proactive. This approach performs well for the heterogeneous networks. That is not
suitable for phone-MANET, if the network does not have support from other resourceful devices. If
the hybrid approach is assumed, the requirement of special node as cluster-head is overcome. As
the cluster based routing protocols depend on the cluster heads and gateways for data packet
transmission, the mall-functioning of these nodes will drop the packets. Due to mobility of nodes,
maintenance of cluster increases and more control overheads are generated. Moreover,
exchanging the responsibility among nodes affect the routing performance as the cluster heads and gateways change frequently due to node mobility. Proper load distribution, reduction of cluster
head responsibility and application of route chasing techniques would provide better performance.
The cluster based protocols generate less overheads than ZRP routing protocols. These protocols
are more scalable than other protocols.

The authors of [24] have proposed a link-state routing protocol that does not depend on other
protocols like ZRP or CBRP. This protocol reduces the control overhead significantly and routing
table size comparatively small. However, the routes can be often broken as the routes are
established over the cluster heads that change frequently due to nodes mobility. The another
reason is that clusters are formed using only one hop neighbours in this protocol. Stable cluster
formation will improve the situation. This protocol is suitable for implementing routing protocol for
resource constrained devices like smartphones. This protocol is not scalable, but scalability can be
improved applying technique of ZRP and managing cluster members to handle different parts of
the zone through the coordination of cluster head. On the other hand, OLSR is quite stable among
the MANET routing protocols. However, the routing table size is bigger than other protocols and
the control overhead increases for the dynamic network. This protocol outperforms than other
protocols when the network size small.

\section{Discussion}
Smartphone based MANET has good prospects to play a vital role in the future wireless
communication systems. However, the deployment of such network in the wider context is not
trivial as the current MANET routing protocols are not directly applicable. The resource constrained
of the smartphones prohibits us to adopt the available routing protocols. The main challenges are
routing table size and protocol control overhead reduction. There are some protocols that can be
modified and combined to design lightweight routing protocols. The combination of clustering and
zone routing can improve the scalability of cluster based routing protocol. Besides, management of
mobility is the crucial factor in MANET protocol designs. Therefore, proper techniques are required
for handeling the mobility that will provide efficient phone-MANET routing protocol.

\bibliographystyle{unsrt}
\bibliography{reflst}

\begin{thebibliography}{10}

\bibitem{roy2010handbook}
Radhika~Ranjan Roy.
\newblock {\em Handbook of mobile ad hoc networks for mobility models}.
\newblock Springer Science \& Business Media, 2010.

\bibitem{bai2010emergency}
Yong Bai, Wencai Du, Zhengxin Ma, Chong Shen, Youling Zhou, and Baodan Chen.
\newblock Emergency communication system by heterogeneous wireless networking.
\newblock In {\em Wireless Communications, Networking and Information Security
  (WCNIS), 2010 IEEE International Conference on}, pages 488--492. IEEE, 2010.

\bibitem{burbank2006key}
Jack~L Burbank, Philip~F Chimento, Brian~K Haberman, and William~T Kasch.
\newblock Key challenges of military tactical networking and the elusive
  promise of manet technology.
\newblock {\em IEEE Communications Magazine}, 44(11), 2006.

\bibitem{adeyeye2011village}
Michael Adeyeye and Paul Gardner-Stephen.
\newblock The village telco project: a reliable and practical wireless mesh
  telephony infrastructure.
\newblock {\em EURASIP Journal on Wireless Communications and Networking},
  2011(1):78, 2011.

\bibitem{gardner2013serval}
Paul Gardner-Stephen, Romana Challans, Jeremy Lakeman, Andrew Bettison, Dione
  Gardner-Stephen, and Matthew Lloyd.
\newblock The serval mesh: A platform for resilient communications in disaster
  \& crisis.
\newblock In {\em Global Humanitarian Technology Conference (GHTC), 2013 IEEE},
  pages 162--166. IEEE, 2013.

\bibitem{jung2007bluetorrent}
Sewook Jung, Uichin Lee, Alexander Chang, Dae-Ki Cho, and Mario Gerla.
\newblock Bluetorrent: Cooperative content sharing for bluetooth users.
\newblock {\em Pervasive and Mobile Computing}, 3(6):609--634, 2007.

\bibitem{camps2013device}
Daniel Camps-Mur, Andres Garcia-Saavedra, and Pablo Serrano.
\newblock Device-to-device communications with wi-fi direct: overview and
  experimentation.
\newblock {\em IEEE wireless communications}, 20(3):96--104, 2013.

\bibitem{zheng2014hybrid}
Chenyu Zheng, Lijun Chen, Douglas Sicker, and Xinying Zeng.
\newblock Hybrid cellular-manets in practice: A microblogging system for smart
  devices in disaster areas.
\newblock In {\em Wireless Communications and Mobile Computing Conference
  (IWCMC), 2014 International}, pages 648--653. IEEE, 2014.

\bibitem{spart}
online.
\newblock https://github.com/projectspan.
\newblock 2014.

\bibitem{chowdhury2013olsr}
Nasim Chowdhury.
\newblock Olsr in android operating system.
\newblock {\em University of Oklahoma, Toronto, Ontario, Canada}, 2013.

\bibitem{hong2002scalable}
Xiaoyan Hong, Kaixin Xu, and Mario Gerla.
\newblock Scalable routing protocols for mobile ad hoc networks.
\newblock {\em IEEE network}, 16(4):11--21, 2002.

\bibitem{neumann2008better}
Axel Neumann, Corinna Aichele, Marek Lindner, and Simon Wunderlich.
\newblock Better approach to mobile ad-hoc networking (batman).
\newblock {\em IETF draft}, pages 1--24, 2008.

\bibitem{sujatha2015uoshr}
Gaini Sujatha and Md~Abdul Azeem.
\newblock Uoshr: Unobservable secure hybrid routing protocol for fast
  transmission in manet.
\newblock In {\em Emerging ICT for Bridging the Future-Proceedings of the 49th
  Annual Convention of the Computer Society of India CSI Volume 2}, pages
  467--474. Springer, 2015.

\bibitem{beijar2002zone}
Nicklas Beijar.
\newblock Zone routing protocol (zrp).
\newblock {\em Networking Laboratory, Helsinki University of Technology,
  Finland}, 9:1--12, 2002.

\bibitem{shaochuan2006aohr}
Wu~Shaochuan, Tan Xuezhi, and Jia Shilou.
\newblock Aohr: Aodv and olsr hybrid routing protocol for mobile ad hoc
  networks.
\newblock In {\em Communications, Circuits and Systems Proceedings, 2006
  International Conference on}, volume~3, pages 1487--1491. IEEE, 2006.

\bibitem{kumar2008genetic}
P~Sateesh Kumar and S~Ramachandram.
\newblock Genetic zone routing protocol.
\newblock {\em Journal of Theoretical \& Applied Information Technology}, 4(9),
  2008.

\bibitem{jiang1999cluster}
Mingliang Jiang.
\newblock " cluster based routing protocol (cbrp)," functional specification.
\newblock {\em IETF Internet Draft, draft-ietf-manet-cbrp-spec-01. txt}, 1999.

\bibitem{baccelli2006olsr}
Emmanuel Baccelli.
\newblock Olsr scaling with hierarchical routing and dynamic tree clustering.
\newblock In {\em IASTED International Conference on Networks and Communication
  Systems}, 2006.

\bibitem{ros2007cluster}
Francisco~J Ros and Pedro~M Ruiz.
\newblock Cluster-based olsr extensions to reduce control overhead in mobile ad
  hoc networks.
\newblock In {\em Proceedings of the 2007 international conference on Wireless
  communications and mobile computing}, pages 202--207. ACM, 2007.

\bibitem{kannhavong2008sa}
Bounpadith Kannhavong, Hidehisa Nakayama, Yoshiaki Nemoto, Nei Kato, and Abbas
  Jamalipour.
\newblock Sa-olsr: Security aware optimized link state routing for mobile ad
  hoc networks.
\newblock In {\em Communications, 2008. ICC'08. IEEE International Conference
  on}, pages 1464--1468. IEEE, 2008.

\bibitem{sharma2013efficient}
Dhirendra~Kumar Sharma, Chiranjeev Kumar, and Surajit Mandal.
\newblock An efficient cluster based routing protocol for manet.
\newblock In {\em Advance Computing Conference (IACC), 2013 IEEE 3rd
  International}, pages 224--229. IEEE, 2013.

\bibitem{huang2009hybrid}
Tsung-Chuan Huang, Kuan-Ping Kho, and Lung Tang.
\newblock Hybrid routing protocol based on the k-hop clustering structure for
  manets.
\newblock In {\em Intelligent Networks and Intelligent Systems, 2009.
  ICINIS'09. Second International Conference on}, pages 197--200. IEEE, 2009.

\bibitem{clausen2003optimized}
Thomas Clausen and Philippe Jacquet.
\newblock Optimized link state routing protocol (olsr).
\newblock Technical report, 2003.

\bibitem{guizani2012new}
Badreddine Guizani, B{\'e}chir Ayeb, and Abder Koukam.
\newblock A new cluster-based link state routing for mobile ad hoc networks.
\newblock In {\em Communications and Information Technology (ICCIT), 2012
  International Conference on}, pages 196--201. IEEE, 2012.

\bibitem{jung2014designing}
Woo-Sung Jung, Hyochun Ahn, and Young-Bae Ko.
\newblock Designing content-centric multi-hop networking over wi-fi direct on
  smartphones.
\newblock In {\em Wireless Communications and Networking Conference (WCNC),
  2014 IEEE}, pages 2934--2939. IEEE, 2014.

\end{thebibliography}
\end{document}